\documentclass{emulateapj}
\newcommand{\xmm}{\textit{XMM--Newton}}
\newcommand{\xte}{\textit{RXTE}}
\newcommand{\chan}{\textit{Chandra}}
\newcommand{\swift}{\textit{Swift}}

\newcommand{\axp}{4U~0142$+$61}

\begin{document}
\title{Long-term X-ray changes in the emission from the \\ anomalous X-ray pulsar 4U~0142$+$61}
\author{M. E. Gonzalez\altaffilmark{1,2}, R. Dib\altaffilmark{1}, V. M. Kaspi\altaffilmark{1,3},
P. M. Woods\altaffilmark{4}, C. R. Tam\altaffilmark{1}, F. P. Gavriil\altaffilmark{5}}

\altaffiltext{1}{Department of Physics, Rutherford Physics Building, 
McGill University, 3600 University Street, Montreal, QC H3A 2T8, Canada.}
\altaffiltext{2}{gonzalez@physics.mcgill.ca; NSERC PGS B.}
\altaffiltext{3}{Canada Research Chair in Observational Astrophysics, Lorne Trottier Chair
in Astrophysics and Cosmology.}
\altaffiltext{4}{Dynetics, 1000 Explorer Blvd. Huntsville, AL 35806, USA; NSSTC, 320 Sparkman 
Drive, Huntsville, AL, 35805, USA}
\altaffiltext{5}{NASA Goddard Space Flight Center, Greenbelt, MD, USA; NPP Fellow, Oak Ridge 
Associated Universities, Oak Ridge, TN}
\begin{abstract}
We present results obtained from X-ray observations of the anomalous X-ray pulsar (AXP)
\axp\ taken between 2000--2007 using \xmm, \chan\ and \swift. In observations taken before
2006, the pulse profile is observed to become more sinusoidal 
and the pulsed fraction increased with time. These results confirm those derived using the 
{\it Rossi X-ray Timing Explorer} and expand the observed evolution to energies below 2 keV. 
The \xmm\ total flux in the 0.5--10 keV band is observed to be nearly constant in observations 
taken before 2006, while an increase of $\sim$10\% is seen afterwards and coincides with the 
burst activity detected from the source in 2006--2007. After these bursts, the evolution 
towards more sinusoidal pulse profiles ceased while the pulsed fraction showed 
a further increase. No evidence for large-scale, long-term changes in the emission as a result of
the bursts is seen. The data also suggest a correlation between the flux and hardness of 
the spectrum, with brighter observations on average having a harder spectrum. As pointed out by 
other authors, we find that the standard blackbody plus 
power-law model does not provide the best spectral fit to the emission from \axp. We also report 
on observations taken with the Gemini telescope after two bursts. These observations show 
source magnitudes consistent with previous measurements. Our results demonstrate the wide 
range of X-ray variability characteristics seen in AXPs and we discuss them in light of current 
emission models for these sources. \\

\end{abstract}


\section{Introduction}
Recent observations of neutron stars have uncovered the wide variety of observational 
manifestations they appear to have, from rotation-powered pulsars and isolated thermally
cooling objects to the so-called ``magnetars" (see, Kaspi, Roberts \& Harding 2006, for a 
review\nocite{krh06}). The latter class includes Anomalous X-ray Pulsars (AXPs) and Soft 
Gamma Repeaters \cite[SGRs; see][for 
a review]{wt06}. Observationally, magnetars exhibit long spin periods of several seconds, 
have persistent X-ray luminosities of $\sim$10$^{34-36}$ ergs s$^{-1}$ and have estimated 
surface dipolar magnetic fields of 0.6$-$7$\times$10$^{14}$ G \cite[see][for a recent review 
of AXPs]{kas07}. Optical and infrared (IR) counterparts have been found for many of these 
objects \cite[e.g.,][]{htv+01,icp+03}. Despite their soft spectrum at low X-ray energies, 
they have also been shown to produce copious amounts of hard X-ray emission 
\cite[e.g.,][]{mcl+04,khdc06}. 

Magnetars are thought to be neutron stars whose X-ray emission is powered by the decay of 
an ultra-high magnetic field ($B>$10$^{15}$~G; Thompson \& Duncan 1995; Thompson \& Duncan 
1996;\nocite{td95,td96a} but see Bhattacharya et al. 2007\nocite{bs07}). The sudden bursts 
of high-energy emission seen from some of these sources are believed to be powered by the 
rearrangement of their magnetic field, while the nature of the optical/IR emission in this 
model is currently under study \citep{bt07}. On the other hand, the active fallback disk 
model argues that the persistent emission at low X-ray energies arises from accretion onto 
the neutron star, while the optical/IR emission is thought to originate from the disk itself 
\citep[e.g.,][]{chn00,alp01}. If only a passive (i.e. non-accreting) disk  is present, it could 
then be responsible for part of the emission at optical/IR wavelengths \cite[][]{wck06}. 
However, a magnetar origin for the high-energy bursts is still needed in all disk models, as 
well as a magnetospheric origin for any pulsed emission at optical/IR wavelengths. The 
large variety of unusual physical phenomena that high magnetic fields can power makes 
magnetars interesting objects to study.

\axp\ is the brightest of all known AXPs. It has a period $P$ = 8.7 s, period derivative $\dot{P}$ =
0.2$\times$10$^{-11}$, inferred surface dipolar magnetic field strength of $B$ = 
3.2$\times$10$^{19}$$\sqrt{P\dot{P}}$ G = 1.3$\times$10$^{14}$ G, and has been detected 
from the mid-IR to hard X-rays \citep{ims94,gk02,dkh+07}. In the mid-IR, \cite{wck06} have found 
evidence for a  passive disk. 
The optical emission was found to be pulsed with a peak-to-peak pulsed fraction of $\sim$29\% 
\citep{km02,dmh+05}. \cite{dv06a} have derived a distance to the source of 3.8$\pm$0.4 kpc. 
In addition, \axp\ has a soft X-ray spectrum that has been fitted traditionally with a 
blackbody plus power-law model with temperature $kT$ $\sim$ 0.4 keV and photon index 
$\Gamma$ $\sim$ 3.3 (e.g., Juett et al. 2002\nocite{jmcs02}; Patel et al. 2003\nocite{pkw+03}). 
However, the extrapolation of this soft X-ray model to the optical/IR overpredicts the observed 
emission if associated with the power-law component, while it underpredicts this emission 
if it is associated with the blackbody component. 

A long-term monitoring campaign for \axp\ has been carried out with the {\it Rossi X-ray Timing 
Explorer} (\xte) for the past 10 yrs. Using these data, \cite{dkg07} reported a slow evolution of 
the pulse profile between 2000 and 2006, as well as a slow increase in the pulsed flux between 
2002 and 2006. These changes may be associated with a possible glitch, or another 
event, that may have occurred between 1998--2000. 
Given the low amplitude of the reported variations, it is important to verify their presence 
with an independent instrument. In addition, the source appears to have entered an active phase 
showing three X-ray bursts in 2006 April--June and 2007 February 
\citep[][]{kdg06,gdkw07a,gdkw07b,gdkw07c} which were followed up by various telescopes. 
The latter of these bursts is the longest and among the most energetic seen from AXPs to date 
\citep{gdkw07c}.

Therefore, constraining the long-term evolution of the emission from \axp\
is important for quantifying all AXP variability phenomena, in the hope of 
discovering correlations that can be checked against models, or uncovering phenomena 
common to all AXPs. Here we present results from observations of this source performed with 
\xmm, \chan\ and \swift, extending from 2000--2007. Indeed, long-term trends in the spectral 
and pulse characteristics are found, confirming the \xte\ results, as well as changes coincident 
with the recent phase of burst activity. Observations in the near-IR after two of these bursts are 
also presented. We find that the overall evolution characteristics support a magnetar origin, 
while the detailed changes suggest that multiple emission mechanisms are likely present. The 
combined effect of these mechanisms is yet to be explored in current modeling of these 
sources. \\

\section{X-ray Observations}

\subsection{\xmm}
\axp\ was observed seven times with \xmm\ between 2002 and early 2007. The details of the observations 
are summarized in Table \ref{tab:obs}. We concentrated on data from the EPIC PN 
instrument (Turner et al. 2001\nocite{taa+01}) as it provided the best combination of longest time 
baseline and highest number of counts, in order to study the long-term evolution of the 
source with high precision.

\begin{table}[h]
\begin{center}
\caption{Observations of \axp  \label{tab:obs}} 
\begin{tabular}{cccc} 
\hline\noalign{\smallskip}
Date & MJD &  CCD Mode/Exp. Time & Counts$^{a}$\\
\hline\noalign{\smallskip}
\multicolumn{4}{l}{\xmm:} \\
13/02/2002 & 52318.3 & Small-Window/2.9 ks & 1.29$\times$10$^{5}$\\
24/01/2003 & 52663.9 & Small-Window/3.8 ks & 1.62$\times$10$^{5}$\\
01/03/2004 & 53065.5 & Timing/29.4 ks & 1.47$\times$10$^{6}$\\
24/07/2004 & 53211.3 & Timing/21.2 ks & 1.07$\times$10$^{6}$\\
28/07/2006 & 53944.8 & Small-Window/3.7 ks & 1.71$\times$10$^{5}$\\
13/01/2007 & 54113.8 & Small-Window/4.4 ks & 1.95$\times$10$^{5}$\\
10/02/2007 & 54141.1 & Timing/8.6 ks & 4.24$\times$10$^{5}$\\
\hline\noalign{\smallskip}
\multicolumn{4}{l}{\chan:} \\
21/05/2000 & 51685.8 & Continuous Clocking/5.9 ks & 1.23$\times$10$^{5}$\\
29/05/2006 & 53915.4 & Continuous Clocking/18.6 ks & 3.86$\times$10$^{5}$\\
10/02/2007 & 54141.3 & Continuous Clocking/20.1 ks & 3.99$\times$10$^{5}$\\
\hline\noalign{\smallskip}
\multicolumn{4}{l}{\swift:} \\
13/02/2005 & 53414.8 & Windowed-Timing/6.6 ks & 2.8$\times$10$^{4}$\\
10/02/2007 & 54141.2 & Windowed-Timing/3.5 ks & 1.6$\times$10$^{4}$\\
\hline\noalign{\smallskip}
\multicolumn{4}{l}{$^{a}$ Net counts in the 0.5--10 keV range. Uncertainties smaller}\\
\multicolumn{4}{l}{than last digit shown.}
\end{tabular}
\end{center}
\end{table}

All \xmm\ observations were reduced and analyzed with SAS v7.0.0 and the latest calibration
files available as of 2007 March. Periods of high particle background were excluded in the 
analysis and standard reduction techniques applied. The PN 
observations performed in the imaging small-window mode had a time resolution of 6 ms, while 
the ones in timing mode had a resolution of 0.03 ms. For the imaging observations, a
circle of 40$''$ radius was used to extract the source counts. For the data in timing mode, the 
source events were extracted from a region 20 pixels wide around the source position. 
Background regions were chosen from regions in the same chip away from the source. The 
total number of background-subtracted counts detected for each observation are
shown in Table \ref{tab:obs}. While the \xmm\ count rates in
each observation are $\sim$45 cts s$^{-1}$, the fast read-out modes in which the instruments
were operated prevented pileup problems.

\subsection{\chan}
\axp\ has been observed three times with \chan\ at sufficiently high time resolution to allow useful 
timing studies (see Table \ref{tab:obs}). The ACIS Continuous Clocking (CC) mode allows for 2.9 ms 
resolution at the expense of one dimension of spatial resolution. The effects of pileup are negligible 
in this mode. The data were reduced using CIAO v3.3.0 and standard 
techniques\footnote{http://cxc.harvard.edu/ciao/threads/aciscctoa/ and 
http://wwwastro.msfc.nasa.gov/xray/ACIS/cctime}. Source events were extracted from a region 4 
pixels wide ($\sim$2$''$) around the peak of the emission, with background regions taken far 
from the source.

\subsection{\swift}
\axp\ has been observed numerous times with \swift. For the purpose of our work, we chose the 
X-ray Telescope (XRT) observations with the highest number of counts and sufficiently high 
time resolution to allow for a study of the timing properties of the pulsar (see Table 
\ref{tab:obs}). The data were reduced applying standard 
screening criteria and using \swift\ Software v2.5a under HEAsoft v6.1.1. Source counts were 
extracted in regions 60 pixels ($\sim$2.4$'$) wide around the peak of the emission, with 
background regions taken far from the source. We only considered events with grade 0 to 
improve spectral resolution and used the latest redistribution matrices (v008). \\

\section{X-ray Pulse Results}\label{sec:tim}
\subsection{Pulse Profiles}\label{ssec:pps}
\begin{figure}
\begin{center}
\includegraphics[width=8.5cm]{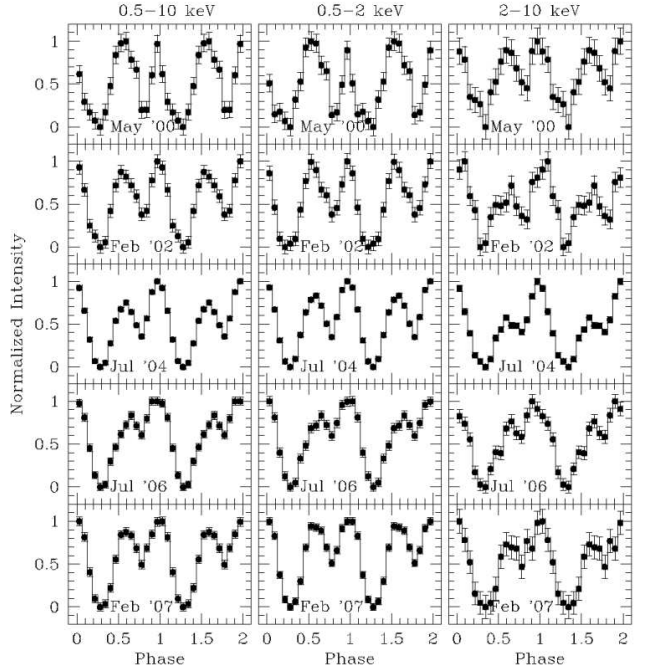}
\caption{\label{fig:pps} Sample pulse profiles for \axp\ in the 0.5--10 keV ({\it left}), 0.5--2 keV 
({\it center}) and 2--10 keV ({\it right}) ranges. The profiles have been normalized to have 
minimum and maximum values between 0 and 1. The top profiles (2000 May) were obtained 
with \chan, and the rest with \xmm. The two bottom profiles were taken after the burst activity
was first detected.}
\end{center}
\end{figure}

The \xmm, \chan, and \swift\ data were used to study the long-term evolution of the X-ray pulse 
profile. The data were transformed to the solar system barycenter and folded at the predicted 
periods for each observation using ephemerides derived from \cite{dkg07}. The data were divided 
into different energy ranges: 0.5--10, 0.5--2 and 2--10 keV. Sample background-subtracted, 
normalized pulse profiles for these bands are shown in Figure \ref{fig:pps}. Prior to 2006, an 
evolution of the pulse profile in the 2--10 keV band is clearly visible and confirms the results 
obtained using \xte\ by \cite{dkg07}. 
In addition, the sensitivity to lower energies allows us to further constrain 
the evolution and conclude that it is also present in the 0.5--2 keV band.

\begin{figure}
\begin{center}
\includegraphics[width=8.2cm]{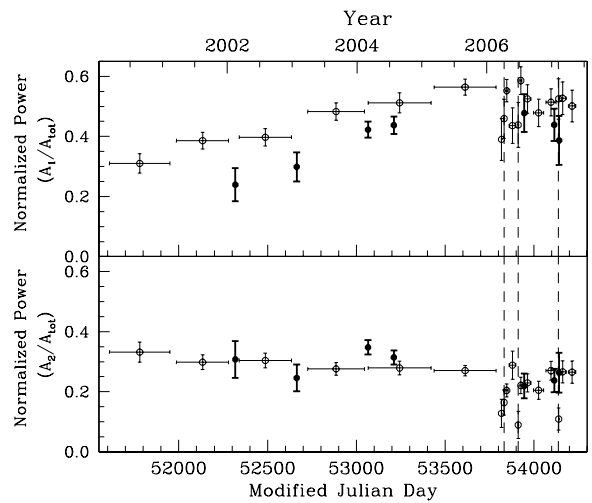}
\caption{\label{fig:harm} Fourier amplitudes of the pulse profiles using \xmm\ 
(bold, solid points) and \xte\ (simple points) in the 2--10~keV band. The pre-burst \xte\ points are 
taken from \cite{dkg07}. The dashed lines indicate the three burst epochs. {\it Top}: Ratio of the 
Fourier amplitude of the fundamental to that of the power in all Fourier amplitudes. {\it Bottom}: 
Ratio of the Fourier amplitude of the second harmonic to that of the power in all Fourier amplitudes.}
\end{center}
\end{figure}

More specifically, before 2006 we find that the relative height of the two peaks increased with time, 
while the depth of the dip between them became less pronounced. This caused the profiles to 
become more sinusoidal, as can be seen from 
a Fourier analysis, in which the ratio of the power in the first harmonic to the total power increased 
while this same ratio for the second harmonic stays fairly constant. Figure \ref{fig:harm} shows these 
ratios for the profiles in the 2--10 keV range obtained from both \xmm\ and \xte\ \citep[see][for 
details]{dkg07}. The ratio of power in the higher harmonics then decreases during this time. 
Given that \xte\ profiles before the bursts were derived by averaging many observations to increase 
the signal-to-noise ratio, the fact that a similar evolution is seen in the individual \xmm\ observations 
confirms the long-term nature of the observed changes. The \chan\ and \swift\ profiles (not shown
in Fig. \ref{fig:harm}) show similar results. After the bursts were detected in 2006, the profiles 
fluctuate more \citep[with more power going to higher harmonics around the time of the bursts, see 
also][]{gdkw07c} and the overall evolution towards more sinusoidal profiles seems to have ceased.

\subsection{Pulsed Fractions}\label{ssec:pfracs}
The fact that the pulse profile evolves with time makes it difficult to determine the pulsed 
fractions (and thus pulsed fluxes) of the source accurately. A few different methods are commonly 
used in the literature to calculate these values; each has advantages and caveats (see Archibald
et al. 2007, in preparation). Here we compare the results obtained from two of these methods: the
root-mean-square (RMS) and area methods. 

We calculate the RMS pulsed fraction using: 
\begin{equation}
PF_{rms} = \frac{\sqrt{2 {\sum_{k=1}^{n}} (({a_k}^2+{b_k}^2)-({\sigma_{a_k}}^2+{\sigma_{b_k}}^2))}}
{\frac{1}{N}{\sum_{i=1}^{N}} {p_i}}
\end{equation}
where $a_k$ is the $k^{\textrm{th}}$ even Fourier component defined as $a_k$ =
$\frac{1}{N} {\sum_{i=1}^{N}} {p_i} \cos {(2\pi k i/N})$, ${\sigma_{a_k}}^2$
is the uncertainty of $a_k$, $b_k$ is the odd $k^{\textrm{th}}$ Fourier
component defined as $b_k$ = $\frac{1}{N} {\sum_{i=1}^{N}} {p_i} \sin {(2\pi k
i/N})$, ${\sigma_{b_k}}^2$ is the uncertainty of $b_k$, $i$ refers to the
phase bin, $N$ is the total number of phase bins, $p_i$ is the count rate in
the $i^{\textrm{th}}$ phase bin of the pulse profile, and $n$ is the maximum
number of Fourier harmonics to be taken into account \citep[we have used 5 harmonics for 
\axp, see also][]{dkg07}. On the other hand, the area pulsed fraction is obtained using:
\begin{equation}
PF_{area} = \frac{\frac{1}{N}{\sum_{i=1}^{N}}({p_i}-{p_{min}})}{\frac{1}{N}{\sum_{i=1}^{N}} {p_i}}
\end{equation}
where $p_{min}$ is the average count rate in the ``minimum" phase 
bins of the profile (as determined by cross-correlating with a high signal-to-noise template). 

\begin{figure}
\begin{center}
\includegraphics[width=8.5cm]{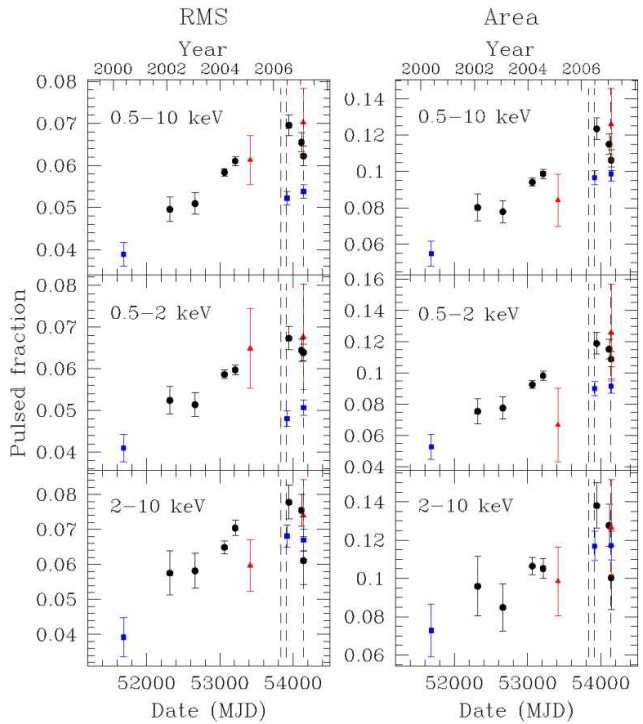}
\caption{\label{fig:pfracs} RMS ({\it left}) and area ({\it right}) pulsed fractions in the 0.5--10 
keV ({\it top}), 0.5--2 keV ({\it center}) and 2--10 keV ({\it bottom}) ranges. Values shown 
correspond to those measured with \xmm\ (black circles), \chan\ (blue squares) and \swift\ (red
triangles). The dashed lines indicate the three burst epochs.}
\end{center}
\end{figure}

While least sensitive to noise, the RMS method returns a pulsed flux number that is affected by 
pulse profile changes. On the other hand, while the area 
method is more physically meaningful, problems in locating the true minimum and its error 
because of noise and binning tend to bias these values upward. Various recommendations have 
been made in order to derive a better estimate for the pulsed fractions for each of these methods 
(Archibald et al. 2007, in preparation) which we have applied here. The resulting values for 
all the observations of \axp\ listed in Table \ref{tab:obs} are shown in Figure \ref{fig:pfracs}. We 
note that the \chan\ values appear to be consistently lower than those found using \xmm\ and
\swift\ and could reflect the calibration uncertainties present in these data (see below).

A significant change in the pulsed fraction over time is seen for both methods. Overall, the pulsed 
fraction has increased with time, reaching an apparent maximum in the observations taken after 
the 2006 burst activity from the source. For example, using the values from the RMS (area) method, 
the pulsed fraction measured with \xmm\ between 2002 and 2006 has increased by 40$\pm$8\%
(58$\pm$12\%), 28$\pm$9\% (57$\pm$15\%) and 35$\pm$14\% (63$\pm$22\%) in the 0.5--10, 
0.5--2 and 2--10 keV bands, respectively. The pulsed fraction also increased significantly in the 
pre-burst observations (2002--2004) and between the observations taken before and after the 
bursts were first detected (2004--2006): in the 0.5--10 keV range, the RMS (area) increase was 
23$\pm$6\% (27$\pm$7\%) and 14$\pm$4\% (25$\pm$7\%) during these time periods, 
respectively. 

In addition, we find evidence for an increase in pulsed fraction with energy in the 
longest \xmm\ observations: in the 0.5--1 keV and 6--10 keV ranges we find RMS values of 
5.1$\pm$0.3\% and 14$\pm$2\%, respectively. However, when using the area method we find
values of 7.9$\pm$0.6\% and 18$\pm$5\% at 0.5--1 keV and 6--10 keV, respectively. Given that 
the pulse profile changes significantly with energy and the area method gives less significant 
changes, we view this suggestive increase in pulsed fraction with energy with caution. \\

\section{X-ray Spectral Results} \label{sec:spec}
The spectra were binned to contain a minimum of 50 counts per spectral 
channel. For the \xmm\ PN data, due to the large number of counts collected from these observations, 
the statistical errors are very
small and a systematic error of 2\% was added to each spectrum (consistent with current calibration 
uncertainties in the PN, Kirsch et al. 2004\nocite{}). The phase-averaged PN spectrum for each 
observation was fitted in the 0.6--10 keV range using the XSPEC package v.11.3.0. Unfortunately, 
we cannot make use of the \chan\ observations taken in CC mode to study the spectral 
characteristics of \axp. While the \xmm\ observations confirm that spectral changes are present in 
the source at the $\sim$10\% level (as will be shown below), the CC data show variations at levels
higher than this, which cannot be corrected for at present due to calibration uncertainties\footnote{See 
CXC Helpdesk ticket \#9114}. The lower signal-to-noise \swift\ XRT data do not contribute to 
constraining the evolution of the spectrum, other than to suggest that changes are present. 
Therefore, the \xmm\ PN was used as the prime instrument to study the long 
term spectral evolution of \axp. 

The \xmm\ PN spectra were fitted simultaneously assuming a common value for the column 
density $N_H$, which was then held constant at its best-fit value. As previously reported, 
single-component models do not describe the emission well and we tried various 
multi-component models, as listed in Table \ref{tab:xmmspec} and shown in Figure 
\ref{fig:specmodels}. 
\cite{rni+07,rtz+07} have 
used various models to fit the emission from \axp\ in the 1--250 keV range. 
\cite{gog07} have also used a spectral model based on a variant of 
the magnetar model to fit the \xmm\ data from \axp. Our fits below are as statistically 
acceptable as those presented by these authors. In addition, although the main focus of 
our paper is to report on the long-term evolution of the emission, we point out that the 
specific values for temperature, emitting area, unabsorbed flux etc. depend heavily 
on the model that is used to describe this emission. This dependence can be seen by 
comparing the parameter values shown in Table \ref{tab:xmmspec} and the unabsorbed 
fluxes shown in Figures \ref{fig:fluxbbpl} and \ref{fig:fluxbb2pl}. We will therefore 
concentrate on flux changes that are seen to be model-independent.

\subsection{Phase-averaged Spectrum}
\begin{figure}[!ht]
\begin{center}
\includegraphics[width=8.2cm]{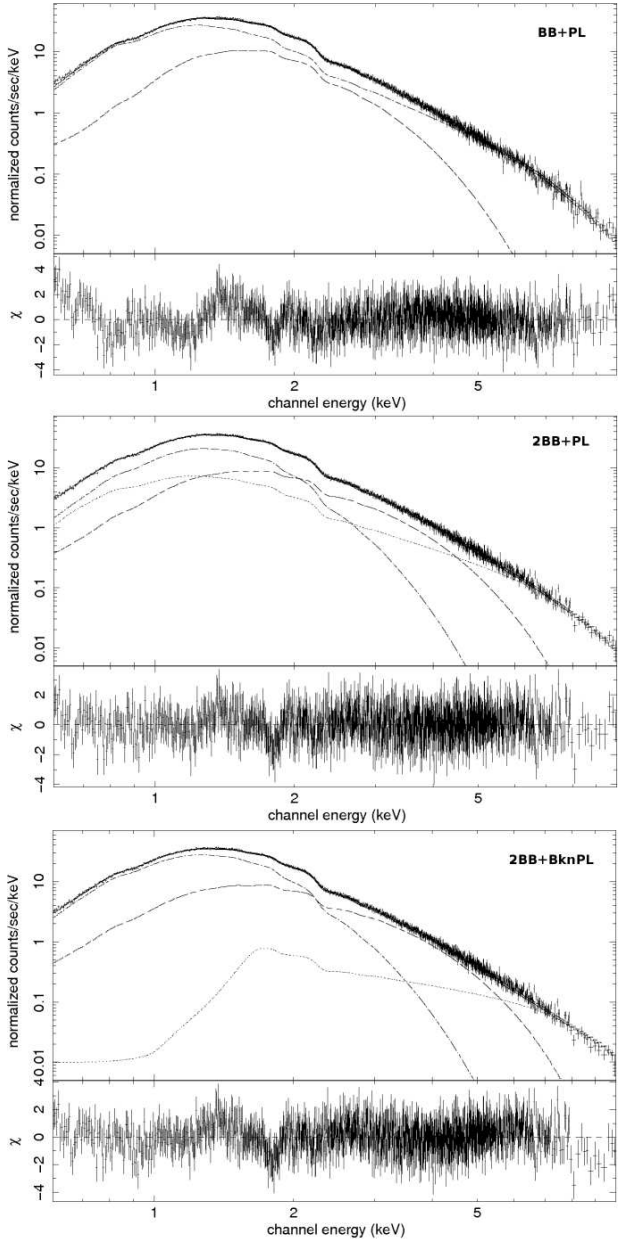}
\caption{\label{fig:specmodels} Best-fit spectral models and residuals obtained for the representative 
2004 March \xmm\ observation of \axp. The models shown are BB+PL ({\it top}), 2BB+PL ($middle$) 
and 2BB+BknPL ({\it bottom}). See Table \ref{tab:xmmspec} for fit values. The individual components 
for each model are also shown. }
\end{center}
\end{figure}

The traditional blackbody plus power-law model used for AXPs (BB+PL) gives results consistent
with those previously reported for the source (e.g., Juett et al. 2002\nocite{jmcs02}, Patel
et al. 2003\nocite{pkw+03}). However, the derived null probability is close to zero 
and many features are evident in the residuals (see Fig. \ref{fig:specmodels}). In addition, 
the derived value for the column density of interstellar absorption for this model is 
inconsistent with that estimated by Durant \& van Kirkwijk (2006\nocite{dv06b}) based 
on the analysis of the high-resolution RGS spectra available from the longer \xmm\ 
observations listed in Table \ref{tab:obs} 
($N_H$~=~(6.4$\pm$0.7)$\times$10$^{21}$ cm$^{-2}$).

\begin{table}[h]
\begin{center}
\caption{Summary of spectral fits to the \xmm\  spectra (see \S\ref{sec:spec} for details) 
\label{tab:xmmspec}} 
\begin{tabular}{lc} 
\hline\noalign{\smallskip}
\hline\noalign{\smallskip}
Parameter\footnote{$kT$ and $R$ represent the observed blackbody temperature and radius,
respectively, while $\Gamma$ is the power-law photon index.} & 
Range of values\footnote{Errors quoted are 1$\sigma$ confidence level. Radii calculated
using a distance of 3.8$\pm$0.4 kpc. Not in chronological order.}\\
\hline\noalign{\smallskip}
\multicolumn{2}{l}{{\it Blackbody+Power-law (BB+PL)}:} \\
\hline\noalign{\smallskip}
$N_H$ & 9.8(2)$\times$10$^{21}$ cm$^{-2}$\\
$kT$ & 0.391(9)--0.44(1) keV\\
$R$ (km) &  5.7(7)--7.3(8) km\\
$\Gamma$ & 3.69(7)--3.771(8)\\
$\chi^2$(dof)& 7407(6843)\\
Probability & 1.1$\times$10$^{-6}$\\
\hline\noalign{\smallskip}
\multicolumn{2}{l}{{\it 2 Blackbodies+Power-law (2BB+PL)}:} \\
\hline\noalign{\smallskip}
$N_H$ & 7.0(2)$\times$10$^{21}$ cm$^{-2}$\\
$kT_{cool}$ & 0.295(9)--0.31(1) keV\\
$R_{cool}$ & 14(2)--16(2) km\\
$kT_{hot}$ & 0.53(1)--0.57(2) keV\\
$R_{hot}$  & 2.9(9)--3.8(6) km \\
$\Gamma$ & 2.76(9)--2.95(8)\\
$\chi^2$(dof)& 6558(6834) \\
Probability & 0.991\\
\hline\noalign{\smallskip}
\multicolumn{2}{l}{{\it 2 Blackbodies+Broken Power-law (2BB+BknPL)}:} \\
\hline\noalign{\smallskip}
$N_H$ & 6.0(1)$\times$10$^{21}$ cm$^{-2}$\\
$kT_{cool}$ & 0.27(1)--0.32(1) keV\\
$R_{cool}$ & 15(1.6)--19(2.5) km\\
$kT_{hot}$ & 0.52(3)--0.61(2) keV\\
$R_{hot}$ & 2.5(3)--4.4(9) km\\
$\Gamma$ & 1.42(5)--1.73(4)\\
$\chi^2$(dof)& 6777(6834)\\
Probability & 0.684\\
\hline\noalign{\smallskip}
\end{tabular}\vspace{-0.3cm}
\end{center}
\end{table}

A two-blackbody model does not fit the observed spectrum well. Instead, a
two-blackbody plus power-law model (2BB+PL) produces the best statistical fit to the data from 
our sample of models. The properties of the two blackbody components could correspond to that 
of a cool component with large emitting area and a hot component with small area. The range of 
temperature and emission area values are consistent with emission from the surface of a neutron 
star. From current magnetar theory, thermal emission from the surface is expected to be scattered 
in the magnetosphere to produce high-energy emission above the thermal peak \citep{tlk02}. 
To simulate this behavior, we have also fit a blackbody plus broken power-law model to the 
data, where the ``break'' energy is set to be the peak of the blackbody model. In addition, the 
power-law component is manually set to contribute negligible emission below this peak 
while having a freely-varying photon 
index above the peak. In this case, we find that a single blackbody model cannot reproduce the 
spectral shape at low energies. Adding another blackbody model significantly improves the fit 
(2BB+BknPL). Plots of these models to the spectrum for the 2004 March observation are shown 
in Figure \ref{fig:specmodels} as an example\footnote{Due to the high statistics, the data show 
residuals at $\sim$1.8 keV that coincide with a Silicon edge in the PN effective area calibration 
(http://xmm.vilspa.esa.es/docs/documents/CAL-TN-0018.pdf)}.

A summary of the range of values for all the observations obtained from these fits is shown in 
Table \ref{tab:xmmspec}. The high quality
data used here expand on what has already been pointed out by other authors: in addition 
to the dubious physical nature of the standard blackbody plus power-law model used to fit
AXP spectra, the data suggest that statistically speaking, it does not reproduce the observed
spectra well, at least for \axp. While we cannot claim that the models presented here are better
physical representations of the observed emission than those presented by, e.g., \cite{gog07},
they describe the observed spectral shape better than a BB+PL and produce reasonable spectral 
parameters.

\begin{figure}
\begin{center}
\includegraphics[width=8.5cm]{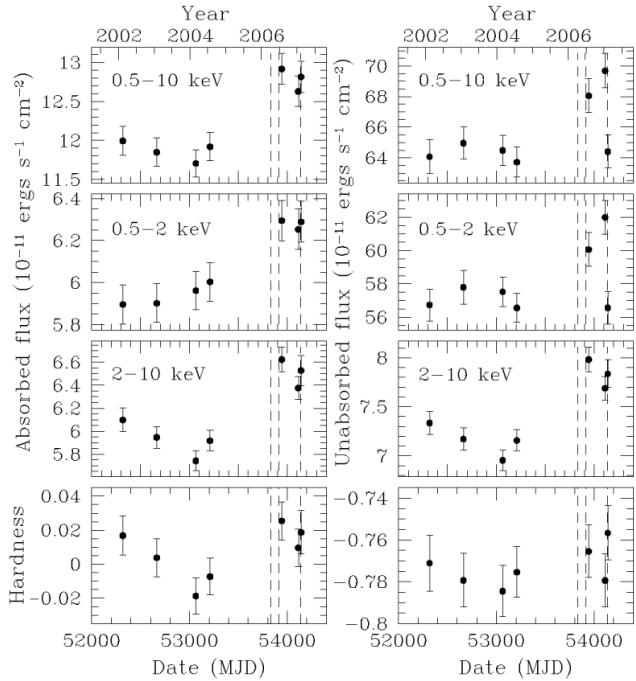}
\caption{\label{fig:fluxbbpl} Phase-averaged fluxes and hardness ratios derived for the BB+PL model.
Absorbed ({\it left}) and unabsorbed ({\it right}) values are shown. The dashed lines indicate the three 
burst epochs.}
\end{center}
\end{figure}

\begin{figure}
\begin{center}
\includegraphics[width=8.5cm]{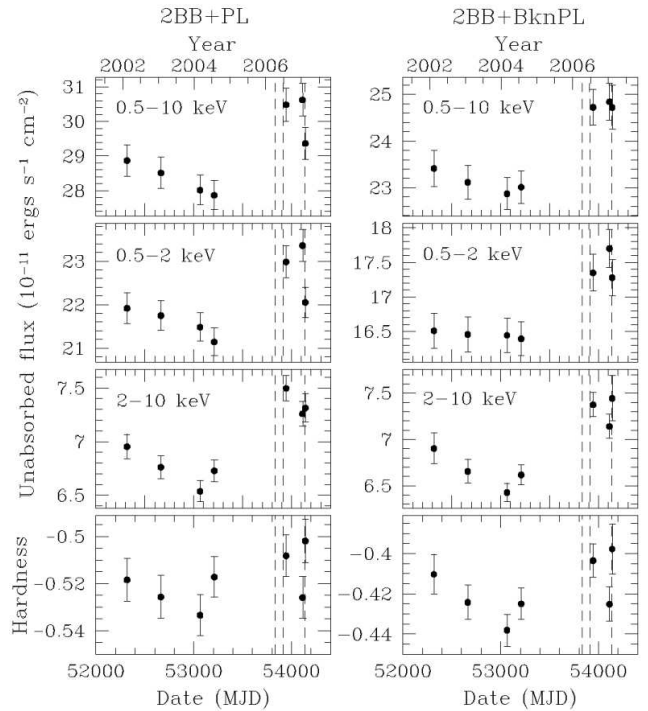}
\caption{\label{fig:fluxbb2pl} Phase-averaged unabsorbed fluxes and hardness ratios derived for the 
2BB+PL ({\it left}) and 2BB+BknPL ({\it right}) model. The dashed lines indicate the three burst epochs.}
\end{center}
\end{figure}

Independent of the model used to fit the data, we find significant changes in the spectral characteristics 
of \axp\ during the span of the observations. Figure \ref{fig:fluxbbpl} shows the absorbed and
unabsorbed fluxes, as well as the hardness of the spectrum, derived using the BB+PL fit. The 
hardness is calculated using $(H-S)$/$(H+S)$, where $S$ is the flux in the 0.5--2 keV range and 
$H$ is the flux in the 2--10 keV range. The values for the unabsorbed flux as given by the 2BB+PL 
and 2BB+BknPL fit are shown in Figure \ref{fig:fluxbb2pl}. Overall, before the recent burst activity, 
the total flux was fairly constant, with a possible decrease in
flux being present and accompanied by an overall softening. After the 2006 burst activity, 
this trend reversed and we find that the 0.5--10 keV flux increased by (10$\pm$3)\%. The increase 
in flux is also energy dependent, with the 0.5--2 and 2--10 keV ranges showing increases of 
(7$\pm$3)\% and (15$\pm$3)\%, respectively. In addition, the spectra for the two 
observations carried out close to burst epochs show evidence of hardening, while the one in 
between shows a softer spectrum. These results are independent of the spectral model used to fit 
the data.

The reported fluxes include a 2\% error due to calibration uncertainties, which greatly 
dominates over statistical errors due to the large number of counts detected. We also 
find that the observed flux variability is mainly caused by changes in the observed PN 
count rate of the source and not uncertainties in the calibration of the instrumental response. 
An almost identical long-term behavior in the count rate and flux is measured 
with MOS1 (operated in timing mode in all but the last observation), albeit with large 
cross-calibration offsets with respect to the PN instrument and smaller number of counts. 
The MOS2 chips were operated in three different modes during 6 of 
these observations (it was turned off on 2002 February), with two imaging observations 
also showing an increase in the flux. Thus, we argue that the observed
spectral changes are intrinsic to \axp\ and not dominated by calibration uncertainties.

\begin{figure}
\begin{center}
\includegraphics[width=8.5cm]{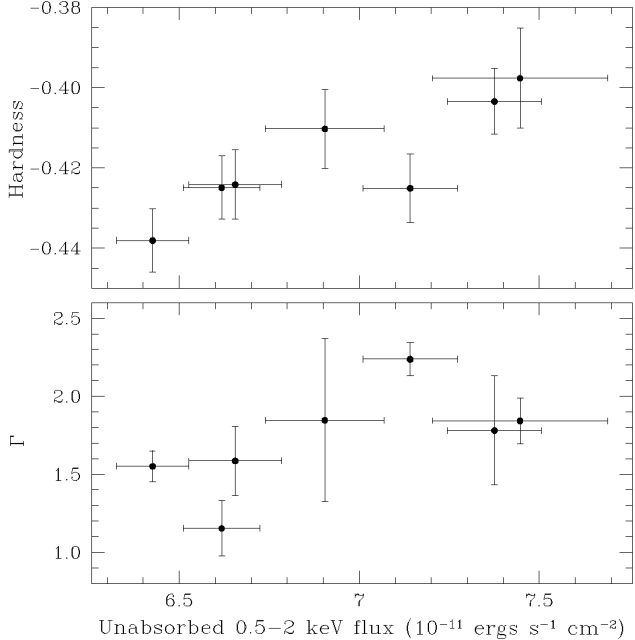}
\caption{\label{fig:hardvsflux} Hardness versus flux from the \xmm\ observations as given by the 
2BB+BknPL model. Absorbed ({\it top}) and unabsorbed ({\it bottom}) fluxes in the 2--10 keV
range are shown. The hardness is derived using $(H-S)$/$(H+S)$, where $H$ is the flux in the 
0.5--2 keV range and $S$ is the flux in the 2--10 keV range.}
\end{center}
\end{figure}

We also find that the hardness of the spectrum is determined mainly by the flux at higher 
energies, and not necessarily by the spectral properties described by model parameters such 
as temperature and photon index (which are also very dependent on the model used to fit the 
spectrum). This is shown in Figure \ref{fig:hardvsflux} 
using the values derived from the 2BB+BknPL model (as it is the most magnetar-inspired of the 
models used), where we find that the hardness versus flux data (top) deviate from a constant at 
the 2.7$\sigma$ level while the photon index (bottom) is consistent with being constant. Similar 
results are obtained for the blackbody temperatures where, if anything, their values are slightly 
lower for observations taken after 2006 when the flux was higher and the spectrum was harder.

\subsection{Pulsed Flux}
\begin{figure}
\begin{center}
\includegraphics[width=8.5cm]{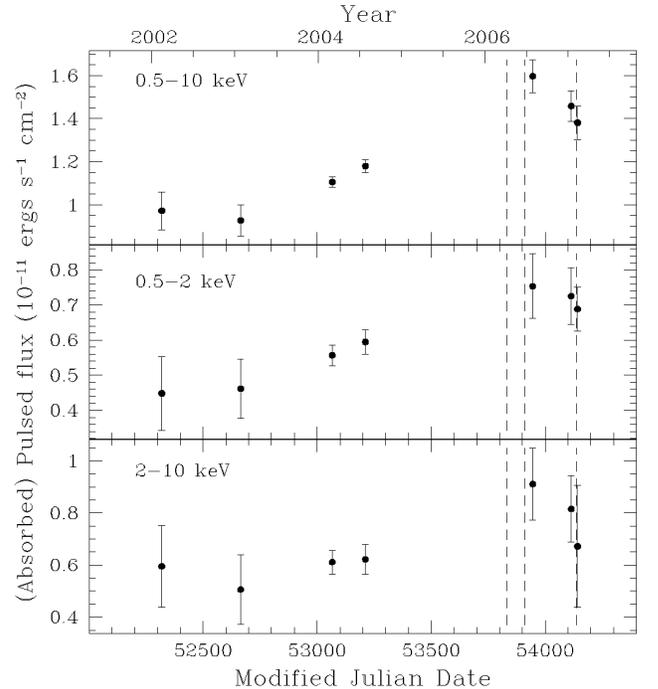}
\caption{\label{fig:pflux} Pulsed fluxes for \axp\ obtained from the absorbed fluxes for the 
2BB+BknPL model and the pulsed fraction derived using the area method. The dashed lines indicate 
the three burst epochs.}
\end{center}
\end{figure}

\begin{figure}
\begin{center}
\includegraphics[width=8cm]{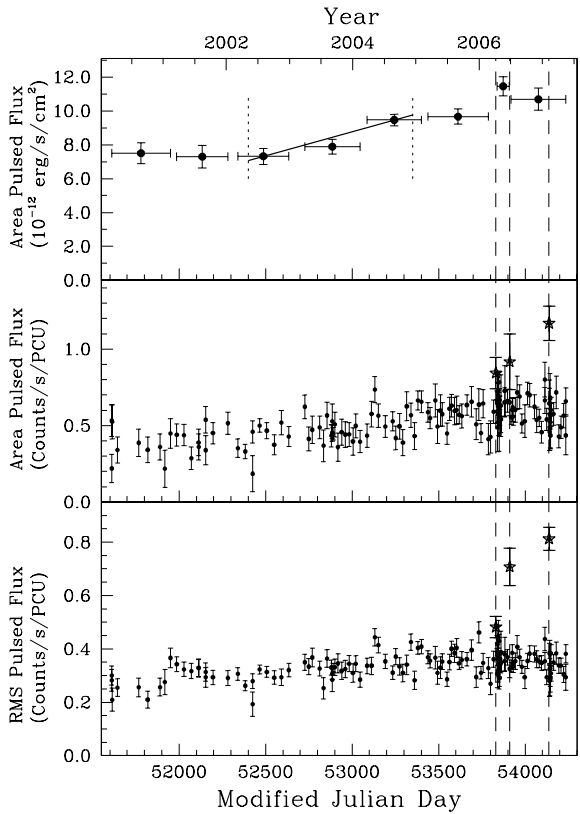}
\caption{\label{fig:pfluxrxte} Pulsed fluxes for \axp\ in the 2--10 keV range obtained using \xte. 
Fluxes in counts derived using the area ({\it center}) and RMS ({\it bottom}) methods are shown for
individual observations (those where bursts were seen are denoted with stars and dashed lines). 
Fluxes in energy
({\it top}) are absorbed and were obtained by combining multiple observations and fitting the 
resulting spectra using XSPEC. The sloped line (bounded by the two short dashed lines) 
shows the $\sim$29\% pulsed flux increase reported in \cite{dkg07}.}
\end{center}
\end{figure}

We also studied the changes in pulsed flux during the observations, the values of which can be
compared to those derived from the \xte\ observations presented by \cite{dkg07}. We have estimated 
the pulsed flux for each observation by taking the phase-averaged flux calculated above and 
multiplying it by the pulsed fraction derived in \S\ref{ssec:pfracs} using the area method. The 
pulsed flux for each observation derived from the 2BB+BknPL model is shown in Figure \ref{fig:pflux} 
(using the absorbed flux, as strictly speaking only an ``absorbed pulsed fraction" can be measured). 
Similarly, in Figure \ref{fig:pfluxrxte} we plot the absorbed pulsed flux in the 2--10 keV range 
derived from 
\xte\ observations. Figure \ref{fig:pfluxrxte} is an updated version of that found in \cite{dkg07} and is
extended to include the most recent observations of \axp. The flux in counts and energy are
both shown, as the former allows for higher time resolution (the observations where bursts were
seen are denoted with stars), while the latter combines multiple observations and allows for direct
comparison with the lower panel of Figure \ref{fig:pflux} \citep[see][for details]{dkg07,gdkw07c}. As 
can be seen from both Figures, the same long-term trend is present in both data sets, albeit \xmm\ 
has larger uncertainties at 2--10 keV due to its lower sensitivity in this energy 
range and the smaller number of observations available. The apparent offset between the \xte\
and \xmm\ fluxes at 2--10 keV is likely caused by cross-calibration uncertainties. \\

\section{Infrared Observations}

\subsection{Gemini}
Two Director's Discretionary Time (DDT) observations of 4U~0142+61
were obtained with the Gemini North Telescope, on 2006 June 30 and 2007
February 13.  Both observations were taken 5--6 days after an X-ray
burst \citep{gdkw07a,gdkw07b}.  $K_S$-band images were made with the
Near-Infrared Imager (NIRI), an ALADDIN InSb 1024$\times$1024 pixel
detector array which, with the f/6 camera, provided a
119.9$\times$119.9 arcsec$^2$ field of view and plate scale of
$0\farcs117$ per pixel.  The standard reduction procedures were
performed using the Gemini package (v1.6) for IRAF (v2.12.2).  Each
frame was a 2$\times$30~s integration; 17 dithered frames in 2006 June 
and 20 in 2007 February, were averaged to make one combined image
for each observation.

\subsection{Results}
The point source 4U~0142+61 was clearly identified in both Gemini observations.
We used DAOPHOT in IRAF for point spread function (PSF) photometry;
the FWHM of the PSF was approximately $0\farcs5$ on 2006 June 30 and
$0\farcs6$ on 2007 February 13.  Using the results of \citet{hvk04},
we calibrated our photometry relatively by measuring the $K_S$-band
magnitude offsets of 10 nearby field stars and applying that offset
to the 4U~0142+61 counterpart, incorporating the offset scatter into
the final uncertainties.  The final calibrated magnitudes are
$K_S = 19.70 \pm 0.05$ and $K_S = 19.86 \pm 0.05$ mag in 2006 June and
2007 February, respectively. The uncertainties are
DAOPHOT-determined and include the calibration uncertainties.

Observations of 4U~0142+61 before the bursts have encompassed a 
large magnitude range, from $K = 19.68 \pm 0.05$ to $K' = 20.78 \pm 0.08$ mag 
\citep{hvk04,dv06c}, consistent with the above values measured after the bursts. 
Therefore, we find no evidence to suggest that the AXP had brightened significantly in 
the near-IR several days after the X-ray bursts. \\

\section{Discussion}
We have found that the X-ray emission from \axp\ changed significantly from 2000--2007.  Before
the 2006 burst activity, the pulse profile became more sinusoidal and the pulsed fraction (and
pulsed flux) increased. Our results agree with those of \cite{dkg07} reported in the 2--10 keV 
range using \xte\ data and we find that these changes are also present in the 0.5--2 keV band. 
During this time, the total flux was approximately constant with time (although a slight decrease 
is suggested depending on the spectral model used). The emission also showed an overall 
softening independent of the assumed spectral model. After 2006, the total flux in the 0.5--10 keV
range increased by $\sim$10\% while the spectrum hardened for those observations close to 
the detected bursts (the spectrum softened in between these observations). The flux increase after
the bursts is also energy-dependent, with higher energies showing a larger increase. During this 
time, the pulse profile evolution towards  more sinusoidal shapes stopped and the pulsed 
fraction was higher than before. We also find that in general, changes in flux and hardness of the 
spectrum appear to be correlated, with observations 
having a higher flux also showing a harder spectrum. This correlation appears to hold at least for 
the small range of flux phase-space that is covered by the current observations. In addition, the 
softening of the spectrum before 2006 agrees with the results presented by \cite{dkg07} and  
the spectral hardening for observations close to detected bursts (in addition to a softening in 
between) agrees with the behavior observed by \cite{gdkw07c}. Our pre-burst absorbed fluxes 
also agree with those of \cite{rni+07} reported using a BB+PL model. 

Anomalous X-ray pulsars exhibit a wide range of behavior in their variability, from sudden energetic
bursts to long-term changes. The AXP 1E 1048--5937 was shown to have large, long-term flares 
of its pulsed flux \citep[one of them lasting about a year,][]{gk04} and variations by a factor of 
$\gtrsim$2 to its phase-averaged flux \citep{tmt+05}. This source has recently become active again,
causing prolonged changes to its observed emission and showing a correlation between hardness
and flux as seen here for \axp \citep{tkg+07}. Changes in the phase-averaged flux of 1RXS 
J170849.0--400910 of $\sim$60\% on a timescale of years have also been reported \citep[][]{cri+07} 
with a correlation between hardness and flux as well
(albeit using various telescopes and instruments). 1RXS J170849.0--400910 was shown to 
have pulse profile changes possibly associated with glitches and low-level pulsed flux variations at 
various epochs, while 1E 1841--045 was shown to have possible long-term pulse profile changes 
and glitches with no obvious radiative changes \citep{dkg07b}. On the other hand, a large 
outburst accompanied by long-term changes in almost all emission characteristics was seen in 
1E 2259+586 that also shows a hardness--intensity correlation\footnote{We note that
a BB+PL model to the observed emission from these sources results in large temperatures and 
steep power-law indices. As such, the power-law component may in fact dominate the observed 
emission below $\sim$1--2 keV, as shown in Figure \ref{fig:specmodels} for \axp. Therefore, 
the hardness--intensity correlations that are measured in terms of the value of the power-law
index may be dominated by the evolution of the low-energy emission from these sources.} 
(Kaspi et al. 2003; Woods et al. 2004; Zhu et al., in preparation) \nocite{kgw+03,wtk+04}.

The very low-level, long-term spectral changes seen here for \axp\ have not been observed thus 
far in other sources and were detected thanks to the high quality of the available data.
The fact that the largest changes are suggested to be accompanied by bursting activity point to a 
common magnetar origin (see below). The overall changes in pulse and spectral properties of 
\axp\ support the view of magnetars as very active sources with a wide range of variability 
characteristics. We now discuss the observed changes
in light of the commonly cited models for AXP emission: the magnetar and disk models. 

\subsection{The Magnetar Model}
In the magnetar model, thermal X-ray emission from the surface provides seed photons which are 
resonantly Compton-scattered (RCS) to higher energies by the enhanced currents in a twisted 
magnetosphere \citep{tlk02,lg06,ft07}. It is also expected that additional thermal emission will be 
produced by return currents from the magnetosphere that heat the surface. In turn, bursts of emission 
are explained as sudden, small-scale reconfigurations of the surface following a crustal yield due to 
a magnetospheric twist. Large outbursts are explained as global reconfigurations and/or reconnections 
of the magnetic field after a large twist. Long-term variability, assuming constant underlying thermal 
emission, is viewed as increases (or decreases) in the twisting of the magnetosphere by currents from 
the stressed crust. The optical depth to scattering increases as the twist angle of the magnetosphere 
increases and in this case we expect a hardening of the spectrum to accompany an increase of the 
emitted flux. This scenario has been used to explain the hardness--intensity correlation observed
in magnetars.
The fact that the predicted correlation between flux and hardness is seen for \axp\ (with 
the brighter observations having a harder spectra) and that the largest changes are observed to 
coincide with a period of increased burst activity support this interpretation.

\cite{og07} have proposed that the hardness--intensity correlation observed in the afterglow
emission from magnetars arises mainly from the cooling crust of the star and less so 
from changes in the magnetospheric currents. For example, most of the burst emission (arising
from a large twist in the magnetosphere) can be deposited 
deep in the crust\footnote{\cite{gogk07} estimate a heating depth of $\sim$2.5 m for the outburst
seen from the transient AXP XTE 1810--197.}, heating
it, and its subsequent cooling dominates the spectral evolution of the star. In this case, the 
observed temperature and total flux would have a direct correlation and could explain the 
hardness--intensity correlation.

However, this does not seem to be the case for \axp. While the latest burst observed from it
was the longest and among the most energetic detected from AXPs thus far \citep{gdkw07b,gdkw07c},
all observations taken after the burst show a rapid return to the previous state without additional 
changes. This suggests that long-term recovery regions (e.g., the inner crust) have not been
significantly affected, or that they were slightly affected and recovered very quickly. In either case, the
bulk of the long-term changes observed after the bursts would be mainly magnetospheric in origin. 
This is supported by the fact that the main spectral changes appear to be dominated by the
emission above 2 keV and that the evolution towards simpler, more sinusoidal profiles appears
to have ceased. 

We also note that the long-term evolution of the pulsed fraction in \axp\ does not show a simple 
correlation with the total flux. This is different from what is observed other sources, such as 
1E 2259+586,  1E 1048.1--5937 and 1RXS J170849.0--400910, where larger phase-averaged 
fluxes correspond to lower pulsed fractions \citep[][]{wtk+04,tmt+05,tkg+07}. The behavior of these 
AXPs could be accounted for, at least in principle, as a growing hot spot on the surface. In the 
transient AXP XTE 
1810--197, as the source slowly fades after a large (undetected) burst  around the end of 2002, 
the pulsed fraction and flux both decrease with time. This behavior may be interpreted as a fading 
hot spot against the background of a large-area cool blackbody \citep{gh07}. 
However, in the case of \axp\ we see a continuous increase in the pulsed fraction independent
of the total flux, suggesting that different mechanisms contribute to this emission with varying
strengths over time. The twisted magnetosphere model generally predicts that 
pulsed fractions should correlate positively with twist angles \citep{ft07}, and this mechanism 
could be responsible for the bulk of the increase during the observed bursting period. However,
the continuous increase before the bursts is still hard to understand.

We then find that the emission from \axp\ shows distinct characteristics from those of 
other AXPs, and while it generally agrees with magnetar models, the complicated evolution of 
the emission characteristics requires more intricate mechanisms than are presently available.

The hard X-ray emission observed in AXPs has also been proposed to arise from the 
twisted magnetosphere, which is thought to act as an accelerator and create a hot corona close 
to the surface of the star \citep{tb05,bt07}. In this case, we would expect the changes seen here 
(mainly those associated with the onset of bursts) to have corresponding changes at hard X-rays.
\cite{dkh+07} report no variability within measurement errors for various observations carried out
before 2006. No measurements after the recent burst activity are reported and 
establishing/constraining any associated variability would be of interest.

The origin of the optical and IR emission in the magnetar model is not well understood, with the 
proposed mechanisms not studied in detail and thus having uncertain correlation with the X-ray flux 
\citep[see][]{tb05}. We find no significant change in the near-IR flux after the bursts that could
be correlated with the overall increase in the X-ray flux during this time. However, given the large 
variability seen from \axp\ in the optical/IR and the subtle nature of the changes in X-rays, we cannot 
test for a possible correlation between the emission at these wavelengths. \\

\subsection{The Disk Models}
The discovery of mid-IR emission from a possible disk around \axp\ has prompted debate as to 
whether it is a passive \citep{wck06} or an active disk \citep{eae+07}. In the case of a 
passive disk the magnetar mechanism is still needed to explain the X-ray emission from the star, 
while an active disk accretes onto a star with a dipole field of $\sim$10$^{12-13}$ G  \citep[a 
magnetar field in the quadrupole or higher components in then needed to explain the bursting 
behavior;][]{eeea07,eae+07}. 
The (unpulsed) optical/IR/UV emission results from the disk as it radiates through viscous energy 
dissipation and by irradiation from the star. Most of the disk radiation, which peaks in the IR, 
comes from the outer regions.

The fact that the main spectral changes are seen to correlate with an increased burst activity 
argue against a disk origin. Although an increased X-ray  flux from the star can affect the 
irradiation emission from the putative disk around \axp, these X-ray changes might or might not 
be accompanied by changes at longer wavelengths depending on the reprocessing efficiency 
of the disk. As the X-ray flux is still observed to be higher in the last observation, changes in the 
optical/IR might 
be present. However, the large range of previously reported IR fluxes do not allow for 
intrinsic changes in the emission of several percent, as observed in the X-ray range, to be readily 
identified. \\

\section{Conclusion}
The observations presented here further demonstrate that variability in AXPs in common. 
The variability takes many forms and proceeds on a wide range of time scales.
The radiative
properties of these objects are seen to vary by orders of magnitude in the case of outbursts and
by a few percent as seen here. The pulse profile and pulsed fraction of \axp\ have undergone 
evident changes during the span of the our observations (2000--2007). Before the bursts were
detected the pulse profile became more sinusoidal, while more complicated changes 
were seen afterwards. On the other hand, the pulsed fraction has increased throughout the 
observations. The total 
flux is observed to have been nearly constant in the observations taken before the bursts, while 
an increase of $\sim$10\% is seen afterwards in the 0.5--10 keV range. The flux increase is energy
dependent, with higher energies showing a larger increase. No evidence for further changes as 
a direct consequence of the bursting activity is seen. The data also
suggest a correlation between flux and hardness of the spectrum, with larger fluxes on average 
having harder spectra. In general, the spectral behavior of the source supports a magnetar origin,
where current models predict that larger twists in the magnetosphere produce 
brighter, harder emission which can coincide with increased burst activity. However, the detailed
evolution of the spectrum and pulse characteristics throughout the observations suggests a more 
complicated scenario, with multiple mechanisms interacting to produce the observed 
properties.
No significant variations in the near-IR emission from the source are detected, consistent with
the few percent change observed in the X-rays flux and uncertainties on how the proposed disk 
around \axp\ would respond to this change.

\acknowledgements
Based on observations obtained at the Gemini Observatory (Program IDs GN-2006A-DD-7 and 
GN-2007A-DD-1), which is operated by the Association of Universities for Research in Astronomy, 
Inc., under a cooperative agreement with the NSF on behalf of the Gemini partnership: the
National Science Foundation (United States), the Particle Physics and Astronomy Research 
Council (United Kingdom), the National Research Council (Canada), CONICYT (Chile), the 
Australian Research Council (Australia), CNPq (Brazil) and CONICET (Argentina). This research 
has made use of data obtained through the High Energy Astrophysics Science Archive Research 
Center Online Service, provided by the NASA/Goddard Space Flight Center. FPG is supported 
by the NASA Postdoctoral Program administered by Oak Ridge Associated Universities at NASA 
Goddard Space Flight Center. This work was also supported by the NSERC Discovery Program,
FQRNT, the Canada Foundation for Innovation, and an R. Howard Webster Fellowship of the 
Canadian Institute for Advanced Research to VMK.

\bibliography{journals1,psrrefs,modrefs}

\end{document}